\documentstyle[preprint,aps]{revtex}
\begin{document}
\preprint{\begin{tabular}{c}
\hbox to\textwidth{May 1997 \hfill YUMS-97-12}\\[-10pt]
\hbox to\textwidth{\hfill SNUTP 97-060}\\[-10pt]
\hbox to\textwidth{\hfill hep-ph/9705420}\\[36pt]
\end{tabular}}
\draft
\title{New Precision Electroweak Tests in Supergravity Models}
\author{Gye T. Park\thanks{E-mail : gtpark@phenom.yonsei.ac.kr} 
and Kang Young Lee\thanks{E-mail : kylee@phenom.yonsei.ac.kr} }
\address {\it Department of Physics, Yonsei University, Seoul, 120-749, Korea}

\maketitle
\begin{abstract}

We update the analysis of the precision electroweak tests in terms of
4 epsilon parameters, $\epsilon_{1,2,3,b}$, 
to obtain more accurate experimental values of 
them by taking into account
the new LEP data released at the 28th ICHEP (1996, Poland).
We also compute 
$\epsilon_1$ and $\epsilon_b$ in the context of the no-scale
$SU(5)\times U(1)$ supergravity model to obtain
the updated constraints by imposing the correlated constraints in terms of
the experimental ellipses in the $\epsilon_1-\epsilon_b$ plane
and also by imposing the new bound on the lightest chargino mass, 
$m_{\chi^\pm_1}\gtrsim 79$ $\,{\rm GeV}$.
Upon imposing these new experimental results, 
we find that 
the situations in the no-scale model are much more favorable than those in the standard model, and 
if $m_t\gtrsim 170$ $\,{\rm GeV}$, then the allowed regions
at the 95\% C.~L. in the no-scale model are
$\tan\beta\gtrsim 4$ and $m_{\chi^\pm_1}\lesssim 120 (82)$ $\,{\rm GeV}$ 
for $\mu>0 (\mu<0)$, which are in fact much more stringent
than in our previous analysis.
Therefore, assuming that $m_t\gtrsim 170$ $\,{\rm GeV}$,
if the lightest chargino mass bound were to be pushed up only
by a few GeV, the sign on the Higgs mixing term $\mu$ in the no-scale model
could 
well be determined from the $\epsilon_1-\epsilon_b$ constraint 
to be positive at the 95\% C.~L.  
At any rate, better accuracy in the measured $m_t$ from the Tevatron 
in the near future combined with the LEP data is most likely 
to provide a decisive test of the no-scale
$SU(5)\times U(1)$ supergravity model. 

\end{abstract}


\section{Introduction}

With the LEP measurements reaching the highest accuracy, 
it has become extremely important for one to perform 
the precision tests of the standard model (SM) and its extensions.
As the top mass, which has long been one of the biggest unknown, 
is being measured much more accurately since its first measurement 
to be now $m_t=175\pm 6$ $\,{\rm GeV}$
from Fermi Laboratory in $\overline{p} p$ collisions \cite{kestenbaum}, 
the standard Higgs mass $m_H$ is the only unknown parameter in the SM.
Therefore
one can investigate the possibilities of revealing 
more about the SM and its extensions of interest.
In the context of supersymmetry, the precision tests can be performed  
within the Minimal Supersymmetric Standard Model (MSSM) \cite{OldEW,BFC,ABC}. 
The problem with such calculations is that there are too many parameters 
in the MSSM and therefore it is not possible to obtain precise predictions 
for the observables of interest.
In the context of supergravity models (SUGRA), 
on the other hand, any observable can be computed in terms of 
at most five parameters: the top-quark mass, the ratio of
Higgs vacuum expectation values ($\tan\beta$), and three universal
soft-supersymmetry-breaking parameters $(m_{1/2},m_0,A)$. This
implies much sharper predictions for the various quantities of interest, as
well as numerous correlations among them. Of even more experimental interest is
$SU(5)\times U(1)$ SUGRA where string-inspired ans\"atze for 
the soft-supersymmetry-breaking parameters allow the theory 
to be described in terms of only three parameters: 
$m_t$, $\tan\beta$, and $m_{\tilde g}$.
Precision electroweak tests in  $SU(5)\times U(1)$ SUGRA 
have been performed in Ref.~\cite{RbKP,LNPZepsb}, 
using the description in terms of the $\epsilon_{1,2,3,b}$ parameters 
introduced in Ref.~\cite{ABC}.
In this paper we update the analysis of the precision electroweak tests 
in terms of these epsilon parameters to obtain 
more accurate experimental values of $\epsilon_{1,2,3,b}$
by taking into account the new data presented at the 28th
ICHEP (July, 1996, Poland) \cite{LEPEWWG} and 
we improve the previous test \cite{RbKP}
by using these values of $\epsilon_{1,2,3,b}$ 
as well as the new bound on the lightest chargino mass \cite{OPAL},
$m_{\chi^\pm_1}\gtrsim 79$ $\,{\rm GeV}$. 

Among $\epsilon_{1,2,3,b}$, $\epsilon_b$ has been of particular interest
because it encodes the $Z\rightarrow b\overline b$ vertex corrections, 
which are mainly proportional to $m_t^2$ and we now know should be
significant due to the heavy top. And therefore it can provide a powerful tool
to constrain $m_t$ indirectly only if $\epsilon_b$ can be determined
experimentally to a good precision.
Along this line, 
$R_b$ ($\equiv{\Gamma(Z\rightarrow b\overline b)
         \over{\Gamma(Z\rightarrow hadrons)}}$), 
which has been measured directly at LEP unlike
$\epsilon_b$ being determined indirectly, has attracted a lot of attentions
because the experimental value for $R_b$ had increased over the years,
resulting in around $3\sigma$ deviation at last above the SM prediction.
In fact, we reported this indication in an attempt to resolve the problem
in the context of the SUGRA \cite{RbKP} and the top-condensate model \cite{kylee}.
However, this so-called ``$R_b$-crisis" became under control to a certain
extent last year when the LEP collaborations announced their new experimental
results on $R_b$ getting much closer to its SM value.
Despite of these new results, the present LEP average on $R_b$ \cite{LEPEWWG},
$R_b=0.2179\pm 0.0012$, still lies around $1.8\sigma$ above its SM value. 
One could certainly
interpret this as a possible manifestation of new physics beyond the SM, 
where at one loop the negative standard top quark contributions are 
cancelled to a certain extent by the contributions from the new particles, 
thereby allowing considerably larger $m_t$ than in the SM.
In fact, the minimal supersymmetric standard model (MSSM) 
realizes this possibility.

In this work, we present the updated analysis of 
the precision electroweak tests in terms of $\epsilon_{1,2,3,b}$ 
by taking into account the latest LEP data
and we explore 
the no-scale $SU(5)\times U(1)$ SUGRA 
in terms of $\epsilon_1$ and $\epsilon_b$
to assess the present status of these models 
being affected by the new LEP data 
and also by the new bound on the lightest chargino mass, 
$m_{\chi^\pm_1}\gtrsim 79$ $\,{\rm GeV}$.

\section{The minimal $SU(5)$ and $SU(5)\times U(1)$ SUGRA models}

The minimal $SU(5)$ \cite{su5sugra} and $SU(5)\times U(1)$ 
\cite{flippedsugra} SUGRA models both contain, 
at low energy, the SM gauge symmetry and the particle content of the MSSM. 
A few crucial differences between the two models are: \\
(i) The unification groups are different, $SU(5)$ versus $SU(5)\times U(1)$.\\ 
(ii) The gauge coupling unification occurs at $\sim 10^{16}$ $\,{\rm GeV}$ 
in the minimal $SU(5)$ model whereas in $SU(5)\times U(1)$ model 
it occurs at the string scale $\sim 10^{18}$ $\,{\rm GeV}$ \cite{LNZstring}.
In $SU(5)\times U(1)$ SUGRA, the gauge unification is delayed 
because of the effects of an additional pair of 
$\bf 10$,$\bf\overline{10}$ vector-like representations 
with intermediate-scale masses. 
The different heavy field content at the unification scale 
leads to different constraints from proton decay. \\
(iii) In the minimal $SU(5)$ SUGRA, proton decay is highly constraining 
whereas it is not in $SU(5)\times U(1)$ SUGRA.

The procedure to restrict 5-dimensional parameter spaces is 
as follows \cite{aspects}.
First, upon sampling a specific choice of ($m_{1/2},m_0,A$) 
at the unification scale and ($m_t,\tan\beta$) at the electroweak scale, 
the renormalization group equations (RGE) are run 
from the unification scale to the electroweak scale, 
where the radiative electroweak breaking condition is imposed 
by minimizing the effective 1-loop Higgs potential, 
which determines the Higgs mixing term $\mu$ up to its sign
\footnote{We define $\mu$ as usual by $W_{\mu}=\mu H_1 H_2$.}.
We also impose consistency constraints such as perturbative unification 
and the naturalness bound of $m_{\tilde g}\lesssim 1\,{\rm TeV}$.
Finally, all the known experimental bounds on the sparticle masses are imposed 
\footnote{We use the following experimental lower bounds 
on the sparticle masses in GeV in the order of 
gluino, squarks, lighter stop, sleptons, and lighter chargino: 
$m_{\tilde g}\gtrsim 150$, $m_{\tilde q}\gtrsim 100$,
$m_{{\tilde{t}}_1}\gtrsim 45$, $m_{\tilde l}\gtrsim 43$, 
$m_{\chi^\pm_1}\gtrsim 45$.}.  
This prodedure yields the restricted parameter spaces for the two models.

Further reduction in the number of input parameters 
in $SU(5)\times U(1)$ SUGRA is made possible 
because in specific string-inspired scenarios for ($m_{1/2},m_0,A$) 
at the unification scale these three parameters are computed 
in terms of just one of them \cite{IL}. 
One obtains $m_0=A=0$ in the {\em no-scale} scenario and 
$m_0=\frac{1}{\sqrt{3}}m_{1/2}$, $A=-m_{1/2}$ in the {\em dilaton} scenario 
\footnote{Note, however, that one loop correction changes this relation 
significantly \cite{cosCKN}.}.

\section{One-loop electroweak radiative corrections and the epsilon
parameters}
There are several schemes to parametrize the electroweak vacuum
polarization corrections  \cite{AB,Kennedy,PT,efflagr}. It can be shown, by
expanding the vacuum polarization tensors to order $q^2$, that one obtains
three independent physical parameters. Alternatively, one can show that upon
symmetry breaking three additional terms appear in the effective lagrangian
 \cite{efflagr}. 
In the $(S,T,U)$ scheme  \cite{PT}, the deviations of the model
predictions from the SM predictions (with fixed SM values for $m_t,m_H$)
are considered as the effects from ``new physics". This scheme is only valid to
the lowest order in $q^2$, and is therefore not applicable to a theory with
light new particles comparable to $M_Z$. 
In the $\epsilon$-scheme \cite{ABC,ABJ}, on the other hand, 
the model predictions are absolute and also valid up to higher
orders in $q^2$, and therefore this scheme is more applicable 
to the electroweak precision tests of the MSSM \cite{BFC} 
and a class of supergravity models \cite{LNPZepsb}.

There are two different $\epsilon$-schemes. 
In the original scheme \cite{ABJ}, $\epsilon_{1,2,3}$ are defined from a basic set of observables 
$\Gamma_{l}, A^{l}_{FB}$ and $M_W/M_Z$.
Due to the large $m_t$-dependent vertex corrections to $\Gamma_b$, 
the $\epsilon_{1,2,3}$ parameters and $\Gamma_b$ can be correlated 
only for a fixed value of $m_t$. 
Therefore, $\Gamma_{tot}$, $\Gamma_{hadron}$ and
$\Gamma_b$ were not included in Ref.~\cite{ABJ}. 
However, in the new $\epsilon$-scheme, 
introduced in Ref.~\cite{ABC}, the above difficulties are overcome 
by introducing a new parameter $\epsilon_b$ to encode
the $Z\rightarrow b\overline b$ vertex corrections. 
The four $\epsilon$'s are now defined from an
enlarged set of $\Gamma_{l}$, $\Gamma_{b}$, 
$A^{l}_{FB}$ and $M_W/M_Z$ without even specifying $m_t$.
This new scheme was adopted in a previous analysis by one of us (G.P.) 
in the context of the $SU(5)\times U(1)$ SUGRA models \cite{LNPZepsb}.
In this work we use this new $\epsilon$-scheme.

With the recent LEP data in Table~I reported by the LEP Electroweak Working Group \cite{LEPEWWG},
we obtain the epsilon variables as follows:
\begin{eqnarray}
\epsilon_1^{exp} &=& (4.0 \pm 1.2) \times 10^{-3}
\nonumber \\
\epsilon_2^{exp} &=& (-4.3 \pm 1.7) \times 10^{-3}
\nonumber \\
\epsilon_3^{exp} &=& (2.3 \pm 1.7) \times 10^{-3}
\nonumber \\
\epsilon_b^{exp} &=& (-1.5 \pm 2.5) \times 10^{-3}~~.
\end{eqnarray}
The lepton universality is assumed for the values of 
$\Gamma_l$ and $A^l_{FB}$ in table 1. 
We note that the $W$ boson mass in table 1 is not directly measured
but only a fitted value like the Higgs mass.
However, only $\epsilon_2$ is related to $m_W$ and
the precision of the $W$ boson mass does not affect
our analysis.

As is well known, the SM contribution to $\epsilon_1$ depends quadratically
on $m_t$ but only logarithmically on the SM Higgs boson mass ($m_H$). 
Therefore upper bounds on $m_t$  have a non-negligible $m_H$
dependence: up to $20$ $\,{\rm GeV}$ stronger when going from a heavy 
($\approx1\,{\rm TeV}$)
to a light ($\approx100$ $\,{\rm GeV}$) Higgs boson. 
It is also known in the MSSM that
the largest supersymmetric contributions to $\epsilon_1$ are expected to
arise from the $\tilde t$-$\tilde b$ sector, and in the limiting case of a very
light stop, the contribution is comparable to that of the $t$-$b$ sector. The
remaining squark, slepton, chargino, neutralino, and Higgs sectors all
typically contribute considerably less. For increasing sparticle masses, the
heavy sector of the theory decouples, and only SM effects  with a {\it light}
Higgs boson survive. However, for a
light chargino ($m_{\chi^\pm_1}\rightarrow{1\over 2}M_Z$), a $Z$-wavefunction
renormalization threshold effect coming from Z-vacuum polarization diagram with
the lighter chargino in the loop can introduce a substantial $q^2$-dependence
in the calculation  \cite{BFC}.
This results in a weaker upper bound on $m_t$ than in the SM.
The complete vacuum polarization contributions from the Higgs sector, 
the supersymmetric chargino-neutralino and sfermion sectors, and also the
corresponding contributions in the SM have been included in our calculations
 \cite{LNPZepsb}. 
However, the supersymmetric contributions to the non-oblique corrections 
except in $\epsilon_b$ have been neglected.

Following Ref.~\cite{ABC}, $\epsilon_b$ is defined from $\Gamma_b$, 
the inclusive partial width for $Z\rightarrow b\overline b$, as
\begin{equation}
\epsilon_b={g^b_A\over{g^l_A}}-1
\end{equation}
where $g^b_A$ $(g^l_A)$ is the axial-vector coupling of $Z$ to $b$ $(l)$.
In the SM, the diagrams for $\epsilon_b$  involve top quarks and
$W^\pm$ bosons  \cite{RbSM}, and the contribution to $\epsilon_b$ depends
quadratically on $m_t$ ($\epsilon_b=-G_F m_t^2/4\sqrt {2}\pi^2 + \cdots$).
In supersymmetric models there are additional diagrams
involving Higgs bosons and supersymmetric particles. The charged Higgs
contributions have been calculated in Refs.~ \cite{Denner,epsb2HD} in
the context of a non-supersymmetric two Higgs doublet model, and the
contributions involving supersymmetric particles in Refs.~ \cite{BF,Rb2HD}.
The main features of the additional supersymmetric contributions are: (i) a negative contribution
from charged Higgs--top exchange which grows as $m^2_t/\tan^{2}\beta$ for
$\tan\beta\ll{m_t\over{m_b}}$; (ii) a positive contribution from chargino-stop
exchange which in this case grows as $m^2_t/\sin^{2}\beta$; and (iii) a
contribution from neutralino(neutral Higgs)--bottom exchange which grows as
$m^2_b\tan^{2}\beta$ and is negligible except for large values of $\tan\beta$
({\it i.e.}, $\tan\beta\gtrsim{m_t\over{m_b}}$) (the contribution (iii) has been
neglected in our analysis).

\section{Results and discussion}
In Fig. 1 we present the results of the calculation of $\epsilon_1$ and $\epsilon_b$ (as described above) for all the allowed points in the no-scale
$SU(5)\times U(1)$ SUGRA for $m_t=170$ $\,{\rm GeV}$ (top row)
and $m_t=180$ $\,{\rm GeV}$ (bottom row).
In the figure we include three ellipses representing the $1\sigma$,
90\% C. L., 95\% C. L. experimental limits obtained from the latest LEP data
as described in the previsous section.
The SM predictions
\footnote{
The SM predictions for $\epsilon$'s are calculated using ZFITTER
\cite{ZFITTER}.
}
for $m_H=100$ $\,{\rm GeV}$ are also shown in Fig. 1
by the symbol ``X'' for $m_t=170, 180$ $\,{\rm GeV}$ as indicated in the figure.
The central values of the new $\epsilon_1^{exp}$ and $\epsilon_b^{exp}$
in Eq.~(1)
have moved from those of the old values toward larger values of
$\epsilon_1$ and smaller values of $\epsilon_b$, and the error bars have also shrunk considerably.
As a consequence, the constraints from $\epsilon_1$ and $\epsilon_b$ 
become much more stringent than before.
In the SM, as can be seen from the figure, one obtains a constraint
on $m_t$, $m_t\lesssim 170$ $\,{\rm GeV}$ at the 95\% C.~L.
where for $m_t$ near the upper bound $m_H$ must be light ($\approx 100$ $\,{\rm GeV}$) due to the logarithmic dependence of $\epsilon_1$ on $m_H$.
However, at the 90\% C.~L., $m_t\lesssim 165$ $\,{\rm GeV}$,
with which the SM can be disfavored by the present measured top mass from the Tevatron. Therefore, in order for the SM to be consistent with the
LEP data in terms of $\epsilon_1^{exp}$ and $\epsilon_b^{exp}$ 
at the 95\% C.~L., $m_t$ must be near the lower end of its measured values.
On the other hand, in the SUGRA, the situations become much more
favorable than in the SM.
From the top row of the Fig.~1, one can see for $m_t=170$ $\,{\rm GeV}$
that there are considerable regions of parameter space to fall inside
the ellipse for the 95\% C.~L.
However, as one can see in the bottom row, for $m_t=180$ $\,{\rm GeV}$  
, only a few points for $\mu <0$ are allowed at the 95\% C.~L.
But these points can be discarded by considering the new bound
on the lightest chargino mass, $m_{\chi^\pm_1}\gtrsim 79$ $\,{\rm GeV}$,
because they simply correspond to $m_{\chi^\pm_1}< 70$ $\,{\rm GeV}$.
Therefore, one obtains
$m_t\lesssim 180$ $\,{\rm GeV}$ at the 95\% C.~L. in the no-scale SUGRA.
For the case in $m_t=170$ $\,{\rm GeV}$, upon imposing the chargino bound
as well, we find in the no-scale SUGRA that there are rather stringent bounds
on the parameter space. For $m_t\gtrsim 170$ $\,{\rm GeV}$, the allowed regions at the 95\% C.~L. are
$\tan\beta\gtrsim 4$ and $m_{\chi^\pm_1}\lesssim 120 (82)$ $\,{\rm GeV}$ 
for $\mu>0 (\mu<0)$.
The tighter constraint for $\mu <0$ comes about because of larger values
of $\epsilon_b$ due to more pronounced chargino-stop contributions
for $\mu <0$.
Therefore, if the lightest chargino mass bound were to be pushed up only
by a few GeV, the sign on $\mu$ in the no-scale SUGRA could well be determined to be positive
from the $\epsilon_1-\epsilon_b$
constraint at the 95\% C.~L.
Now, as to the constraints in the no-scale model at the 90\% C.~L.,
on the other hand, one can see that there are some points for $\mu <0$
falling inside the 90\% C.~L. ellipse, which are in fact excluded again
by the chargino mass bound. And hence, in the no-scale SUGRA,
one can conclude that $m_t\lesssim 170$ $\,{\rm GeV}$ remains allowed
at the 90\% C.~L.
, which implies that the present LEP data prefers the SUGRA rather than
the SM and also the top mass near the lower end of its current measured values
at the Tevatron. 
The major features of the constraints from $\epsilon_1$ and $\epsilon_b$
for the no-scale SUGRA are summarized in the Table~II.

\section{Conclusions}
We have updated the analysis of the precision electroweak tests in terms of
4 $\epsilon$ parameters, $\epsilon_{1,2,3,b}$, to obtain more accurate experimental values of them by taking into account
the new LEP data released at the 28th ICHEP (1996, Poland).
We have also computed the one-loop electroweak corrections in the form of
$\epsilon_1$ and $\epsilon_b$ in the context of the no-scale
$SU(5)\times U(1)$ supergravity model to obtain
the updated constraints by imposing the correlated constraints in terms of
the experimental ellipses in the $\epsilon_1-\epsilon_b$ plane
as well as the new bound on the lightest chargino mass, 
$m_{\chi^\pm_1}\gtrsim 79$ $\,{\rm GeV}$.
Upon imposing these new experimental ellipses, one obtains in the SM 
perilously low bounds on $m_t$,
$m_t\lesssim 170 (165)$ $\,{\rm GeV}$ at the 95\% (90\%) C.~L.
However, the situations become much more favorable
in the no-scale SUGRA, and we find in the model that
$m_t\lesssim 180 (170)$ $\,{\rm GeV}$ remains allowed at the 95\% (90\%) C.~L.
and 
if $m_t\gtrsim 170$ $\,{\rm GeV}$, then the allowed regions
at the 95\% C.~L. are
$\tan\beta\gtrsim 4$ and $m_{\chi^\pm_1}\lesssim 120 (82)$ $\,{\rm GeV}$ 
for $\mu>0 (\mu<0)$, which are in fact much more stringent
than in our previous analysis.
Therefore, assuming that $m_t\gtrsim 170$ $\,{\rm GeV}$,
if the lightest chargino mass bound were to be pushed up only
by a few GeV, the sign on $\mu$ in the no-scale SUGRA 
could well be determined to be positive from the $\epsilon_1-\epsilon_b$
constraint at the 95\% C.~L.  
Moreover, should the central value of the measured $m_t$ in the near future remain as it is now, the $\epsilon_1-\epsilon_b$ constraint would prefer the no-scale model at the 95\% C.~L.
At any rate, better accuracy in the measured $m_t$ from the Tevatron in the near future combined with the LEP data
is most likely to provide a decisive test of the no-scale
$SU(5)\times U(1)$ supergravity model. 

\acknowledgments
The work of G.P. has been
supported in part by the Korea Science and Engineering Foundation(KOSEF) Grant
No. 971-0201-006-2 and in part by the KOSEF through the SRC program of SNU-CTP.
The work of K. Y. L. has been
supported by Korea Research Foundation
(KRF) and Research University Fund of College of Science
at Yonsei University by Ministry of Education (MOE) of Korea.

\begin{figure}
\caption{The correlated predictions for the $\epsilon_1$ and $\epsilon_b$
parameters in units of $10^{-3}$ in the no-scale $SU(5)\times U(1)$ 
SUGRA for $m_t=170$ $\,{\rm GeV}$ (top row) and
for $m_t=180$ $\,{\rm GeV}$ (bottom row).
The contours for $1\sigma$, 90\% C.L., and 95\% C.L. obtained
from the LEP data announced at the ICHEP (1996, Poland) are included.
For comparison, the SM predictions for $m_H=100$ $\,{\rm GeV}$ are also shown
by the symbol ``X'' for the values of $m_t$ as indicated.}
\label{epsb1}
\end{figure}
\begin{figure}
\end{figure}
\begin{figure}
\end{figure}

%
%

\begin{table}
\begin{center}
\caption{
The LEP data reported by the LEP Electroweak Working Group 
at the 28th ICHEP (1996, Poland).
}
\label{Table1}
\vspace{2cm}
\begin{tabular}{|clc|ccc|}
 & $M_W$ & & & 80.2780  $\pm$  0.0490 GeV& \\
 & $M_Z$ & & & 91.1863  $\pm$  0.0020 GeV& \\
 & $\Gamma_l$ & & & 83.91  $\pm$  0.11 MeV& \\
 & $A_{FB}^l$ & & & 0.0174  $\pm$  0.0010 & \\
 & $\Gamma_b$ & & & 379.9  $\pm$  2.2 MeV& \\
\end{tabular}

\end{center}
\end{table}

\vspace{2cm}

\begin{table}
\caption{
The major features of the constraints from $\epsilon_1$ and $\epsilon_b$ 
for the no-scale SUGRA.
}
\label{Table2}

\vskip 2.0cm
\begin{tabular}{|ccc|cc|}
&          &&no-scale $SU(5)\times U(1)$&\\ 
\hline
&           &&$m_t\lesssim 180$ $\,{\rm GeV}$ for any $\tan\beta$&\\
&(95\% C.L.)&&For $m_t\gtrsim 170$ $\,{\rm GeV}$, $\tan\beta\gtrsim 4$ and&\\
&          && $m_{\chi^\pm_1}\lesssim 120$ (82) $\,{\rm GeV}$ for $\mu>0$ ($\mu<0$)&\\ 
\hline
&(90\% C.L.)&&$m_t\lesssim 170$ $\,{\rm GeV}$ for any $\tan\beta$&\\
\end{tabular}
\end{table}

\end{document}